\documentclass[12pt]{iopart}
\begin{document}
\jl{1}

\letter{ The Dual Phases of Massless/Massive Kalb--Ramond Fields}

\author{Anais Smailagic\dag\footnote[3]{e-mail
		address: \texttt{anais@etfos.hr}},
		Euro Spallucci\dag\footnote[5]{e-mail
		address: \texttt{spallucci@trieste.infn.it}}}
\address{\dag\ Department of Physics, Faculty of Electrical Enginnering,
University of Osijek, Croatia}
\address{\ddag\ Dipartimento di Fisica Teorica
Universit\`a di Trieste, and INFN, Sezione di Trieste}

\begin{abstract}
  We have developed dualization of ordinary and  ``Stueckelberg compensated'' 
  massive phase for the Kalb--Ramond theory. The compensated phase allows
  study of the interplay between spin jumping and duality. We show that spin 
  jumping is caused by mass, while gauge symmetry is {\it not} necessary 
  for this effect to take place.
  \end{abstract}
    
\pacno{11.15}

\submitted

\maketitle

\noindent
    
   Recently,  we have described   a general procedure for
   dualizing {\it massless} $p$-forms in $D$ dimensions, including 
   non--trivial limiting cases $p+1=D$ \cite{aass},\cite{pr}. The dualizing 
   procedure has been further successfully extended to  non--Abelian  
   Kalb--Ramond (KR) field theory \cite{nonab},\cite{lan}. Summary of the 
   results is that massless $p$--forms are dualized to $D-p-2$ forms according 
   to the scheme:
   
   \begin{equation}
    m=0\quad: p\longleftrightarrow D-p-2
    \label{m1}
    \end{equation}
   
   where the number of dynamical degrees of freedom is given by $ D-2\choose p$.
   On the other hand, the number of dynamical degrees of freedom for
   the same rank, but {\it massive} $p$--form,  is given by $ D-1\choose p$.
   Thus, simple counting shows that massive $p$--forms cannot be dual to
   the same fields as in the massless case. In fact, they 
   have to be dual to a  $D-p-1$--form as follows 
   
   \begin{equation}
    m> 0\quad: p\longleftrightarrow D-p-1
    \label{m2}
    \end{equation}
   
   As a consequence of different schemes, the same $p$--form, can describe
   fields of different spin in massless and massive phases. This phenomenon
   has been already observed in \cite{at} and 
   is known as {\it spin jumping}. From the  dualization schemes (\ref{m1})
   and (\ref{m2}) spin jumping turns out to be a general
   property of  $p > 1 $ forms. In $D=4$ dimensions spin jumping takes place
   for KR ($p=2$) and $p=3$--field. The latter corresponds to the limiting
   case $p=D-1$ and has special properties in the massless phase as described 
   in \cite{aass}, while the former clearly exhibits spin jumping in both
   phases. Furthermore, KR field theory has also been widely investigated as 
   an alternative to the Higgs
   mechanism \cite{at}, \cite{orland}, as well as in relation to the problem 
   of confinement in $QCD$ and strong/weak coupling duality \cite{savit}. 
   More recently, the generalization of the Higgs/Stueckelberg mechanism for 
   higher dimensional branes has been also discussed in the path integral 
   formalism \cite{qdt},\cite{qt},\cite{luca}.\\
   
   For the above mentioned reasons,
   in this letter we shall describe the extension of the dualization
   procedure for massless $p$--forms to the 
   case of {\it massive} KR fields,  KR field being 
   the lowest rank $p$--form exhibiting the spin jumping.\\
   We shall start by reviewing the 
   dualization of  massless  KR field following \cite{aass}.\\
   One starts by introducing a parent action,  which in this letter we
   choose to be of the first order.
   What we mean is an action described in terms of independent  $3$--form 
   field  $H$ and KR gauge potential $B$:
       
     \begin{equation}
     S[\, H\ , B\,]=\int d^4x \left[\,
      {1\over 12} H_{\lambda\mu\nu} \, H^{\lambda\mu\nu} 
     -{1\over 6}H^{\lambda\mu\nu}\partial_{[\, \lambda} B_{\mu\nu\,]}\, \right]
     \label{uno}
     \end{equation}
       
     We shall apply the dualization procedure in a path integral framework, 
     thus we define the generating functional as 
     
     \begin{equation}
     Z[\, J\,]= \int [DH]\, [DB]\, \exp\left\{ - S[\, H\ , B\,]- {g\over 2}
     \int d^4x \, B_{\mu\nu}\, J^{\mu\nu}\,\right\}
     \end{equation}
    
     The above parent generating functional is invariant under KR gauge 
     symmetry 
     
     \begin{equation}
     \delta H_{\lambda\mu\nu}=0\ ,\quad
	\delta B_{\mu\nu}=\partial_{[\, \mu}\Lambda_{\nu\,]}\ ,\quad
     \end{equation}
     
     provided we define an appropriate invariant integration measure and
     the external current $J^{\mu\nu}$ is chosen to be divergence free.
     \\
     The reason we have adopted path integral formulation is that,
     in case of an Abelian theory, it gives 
     identical results as the algebraic approach \cite{aass} and also allows 
     one to prove the quantum equivalence of dual theories \cite{jap}.
     Furthermore, it is the only formalism that allows non-Abelian extension
     of the dualization procedure \cite{nonab}.
     The dualization proceeds by integrating  the $B$ field which gives
     a delta--function:

     \begin{equation}
     \fl \int[DB]\,\exp\left\{- {1\over 2}
     \int d^4x \, B_{\mu\nu}\,\left(\, \partial_\lambda H^{\lambda\mu\nu}-g\,
      J^{\mu\nu}\,\right)\, \right\}= \delta\left[\, \partial_\lambda
      H^{\lambda\mu\nu}-g\, J^{\mu\nu}\,\right]
      \end{equation}

     The delta--function  imposes classical equations of motion \cite{euroj} 
     on $H$, which can be solved as
       
     \begin{equation}
     H^{\lambda\mu\nu}= \epsilon^{\lambda\mu\nu\rho}\partial_\rho \phi
      + g\, \partial^{[\,\lambda}\, {1\over\partial^2 }\, J^{\mu\nu\,]}\ .
     \label{hscalar}
     \end{equation}
       
      Secondly, integrating out the $H$ field, using the
      solution (\ref{hscalar}), one finds the   generating functional
      in terms of  massless, scalar, field $\phi$ which is the dual of the
      KR potential $B$:
      
      \begin{eqnarray}
     Z[\, J\,] =  \int [D\phi]\, 
     \exp\left\{ - {1\over 12}
     \int d^4x \,\left(\,  \epsilon^{\lambda\mu\nu\rho}\partial_\rho \phi
      + g\, \partial^{[\,\lambda}\, {1\over\partial^2 }\, J^{\mu\nu\,]} \,
      \right)^2 \, \right\}\nonumber\\
       =  \int [D\phi]\, 
     \exp\left\{ - {1\over 2}
     \int d^4x \,\left[\,\partial_\mu \phi \, \partial^\mu \phi 
      - {g^2\over 2}\, J^{\mu\nu}\, {1\over\partial^2 }\, J_{\mu\nu}\,\right]\,
      \right\}\ . 
      \label{effscal}
      \end{eqnarray}
       
       This concludes a brief review of the known duality between a massless KR
       field and a scalar field using first order formalism.\\
       On the other hand,  {\it massive} KR field 
      has a different number of dynamical degrees of freedom from the massless
      case. Thus, massive KR field cannot be dual to a scalar field. 
       To  dualize a massive KR field we shall start from the parent action 
       
       \begin{equation}
     S[\, H\ , B\,]= \int d^4x\left[\,
     {1\over 12} H_{\lambda\mu\nu} \, H^{\lambda\mu\nu} 
     -{1\over 6}H^{\lambda\mu\nu}\partial_{[\, \lambda} B_{\mu\nu\,]}
      +{m^2\over 4} B_{\mu\nu}\,  B^{\mu\nu}  \,\right]
     \end{equation}
     
     The KR gauge invariance is explicitly broken and
     there is no a priori reason why should the external current $J^{\mu\nu}$ 
     be divergence free. However, we shall still impose $\partial_\mu 
J^{\mu\nu}=0$
     in order to confine gauge symmetry breaking only within the mass term.
     We proceed  by integrating out the $B$ field:
     
     \begin{eqnarray}
    \fl Z[J] =  \int [DH] [DB]\,\nonumber\\
    \fl  \times \exp\left\{ \int d^4x \,\left[{1\over 12}
     H_{\lambda\mu\nu} \, H^{\lambda\mu\nu}  - {1\over 2}
     B_{\mu\nu}\,\left(\, \partial_\lambda H^{\lambda\mu\nu}-g\,
      J^{\mu\nu}\,\right)  +{m^2\over 4} B_{\mu\nu}\,  B^{\mu\nu}  \,\right] 
      \right\} \nonumber\\
    \fl   = \int [DH] \exp\left\{\, \int d^4x\, \left[\,
      {1\over 12} H_{\lambda\mu\nu} \, H^{\lambda\mu\nu} +{1\over 4m^2 }
      \left(\, \partial_\lambda H^{\lambda\mu\nu}-g\, J^{\mu\nu}\,\right)^2
      \,\right]
      \right\}\label{h}
      \end{eqnarray}
    
     In order to obtain dual theory 
     we express $H$ in (\ref{h}) through its  Hodge Dual \footnote{
     It is also possible to formulate second order parent Lagrangian
     for the massive KR field, which is the extension of the procedure
     in \cite{aass}.  It gives the same result as the above shortcut
     procedure.} as
      
      \begin{equation}
      H^{\lambda\mu\nu}\equiv m\,\epsilon^{\lambda\mu\nu\rho}\, A_\rho\ .
      \label{hodge}
      \end{equation}
      
      Inserting (\ref{hodge}) in (\ref{h}) gives the dual action 
      of a massive vector field

      \begin{eqnarray}
     \fl S[\, A\ , J\,]  =  \int d^4x\, \left[\, {1\over 4 }
      \left(\, \epsilon^{\lambda\mu\nu\rho}\,\partial_\lambda \, A_\rho 
       -{g\over m}\, J^{\mu\nu}\,\right)^2 + {m^2\over 2 }\, A_\mu\,  A^\mu
      \,\right]\nonumber\\
     \fl  \equiv   \int d^4x\, \left[\, {1\over 4 }\, F^{\mu\nu}(A)\, 
F_{\mu\nu}(A)
     -{g\over 6m}\, 
       \, J^{\ast\, \mu}\, A_\mu  + {m^2\over 2 }\, A_\mu\,  A^\mu+ 
       {g^2\over 4m^2}\, J_{\mu\nu}\, J^{\mu\nu}
      \,\right] \nonumber\\
       \label{effm}
       \end{eqnarray}

      where, we have  introduced  the  current  $ J^{\ast\, \rho}\equiv
     \epsilon^{\lambda\mu\nu\rho}\,\partial_{[\,\lambda} \, J_{\mu\nu\,]}$.
       Thus,  we have shown that the massive KR field is
	dualized to a  vector field of the Proca type as expected from
	the scheme (\ref{m2}). Both 
	fields have {\it three} physical degrees of freedom. Results 
	(\ref{effscal}) and (\ref{effm}) display the effect of spin 
	jumping of the  KR field from  spin zero to spin one.\\
	At this point, we would like to develop a dualization procedure
	for a {\it massive, but gauge invariant, } KR theory. 
	An effective way to achieve this goal is to apply a Stueckelberg
	compensation procedure \cite{pr}. In this way,
	one can combine two properties, mass and gauge symmetry,
	which were mutually exclusive in (\ref{effscal}) and (\ref{effm}).
	The reason for considering compensated theory is our intention
	to determine whether spin jumping is caused by the presence of mass
	or by gauge symmetry. For the KR field Stueckelberg 
	compensation amounts to the following substitution
	$B_{\mu\nu}\longrightarrow B_{\mu\nu} -\partial_{[\, \mu}\phi_{\nu\,]}$
	where $\phi_\nu$ is the compensating vector field transforming as
	$\delta \phi_\nu= \Lambda_\nu$ under KR symmetry.
	 The Stueckelberg parent action turns out to be:
	
	\begin{eqnarray}
    \fl S[\, H\ , B\ ,\phi\,]=  \int d^4x\left[\, 
     {1\over 12}\, H_{\lambda\mu\nu} \, H^{\lambda\mu\nu} 
     -{1\over 6}\, H^{\lambda\mu\nu}\, \partial_{[\, \lambda}\, B_{\mu\nu\,]}
      +{m^2\over 4}\,\left(\, 
       B_{\mu\nu} -\partial_{[\, \mu}\phi_{\nu\,]} \,\right)^2 \right.
       \nonumber\\
   \fl -{g\over 2}\left. \left(\, 
       B_{\mu\nu} -\,\partial_{[\, \mu}\phi_{\nu\,]} \,\right) \,
       J^{\mu\nu}
       \,\right]\label{hbfi}
      \end{eqnarray} 
	
	As in the previous cases, we shall choose a divergence free external
	current $J^{\mu\nu}$, thus the compensator in the last term of
	(\ref{hbfi}) drops out.
	Dualization starts by integrating out the compensator.
	In order to simplify integration of the Stueckelberg
	field, let us linearize the mass term introducing  an
	additional field $C_{\mu\nu}$  which transforms 
	the parent action   (\ref{hbfi})    into 

      \begin{eqnarray}
     \fl  S[\, H\ , B\ ,\phi\ , C\,] =  \int d^4x\left[\,
     {1\over 12} H_{\lambda\mu\nu} \, H^{\lambda\mu\nu} 
     -{1\over 6}H^{\lambda\mu\nu}\partial_{[\, \lambda} B_{\mu\nu\,]}
      -{1\over 4}C^{\mu\nu}  C_{\mu\nu}\right.  \nonumber\\
      \fl +  m\, C^{\mu\nu}\,\left.  \left(\, 
       B_{\mu\nu} -\partial_{[\, \mu}\phi_{\nu\,]} \,\right)\right.
    \left.  -{g\over 2} B_{\mu\nu} \, J^{\mu\nu}\,\right]
      \end{eqnarray}

	Now  the compensator $\phi_\mu$ is a Lagrange multiplier and its 
	integration gives  a delta function as 

	\begin{equation}
       \int [D\phi]\exp\left\{\, \int d^4x\, m\, \phi_\nu \, \partial_\mu\,
       C^{\mu\nu} \,\right\}=
       \delta\left[\, \partial_\mu C^{\mu\nu} \,\right]
      \end{equation}

	The above delta function leads to the solution for the field $C$ as

	\begin{equation}
	 C^{\mu\nu}={1\over 2}\,
       \epsilon^{\mu\nu\rho\sigma}\,\partial_{[\,\rho}A_{\sigma\,]} \label{ccl}
	\end{equation}

	Subsequent integration over the  $B$ field gives additional delta 
	function

	\begin{eqnarray}
       \int [D B]  \exp\left\{\, {1\over 2}\int d^4x\, B_{\mu\nu}\right.  
      \left. \left(\, \partial_\lambda
       H^{\lambda\mu\nu} + m\, C^{\mu\nu}   -g\, J^{\mu\nu}\,\right)\,
       \right\}=
       \nonumber\\
        \delta\left[\, \partial_\lambda
       H^{\lambda\mu\nu} + m\, C^{\mu\nu}   -g\, J^{\mu\nu}\,\right]\ ,
      \end{eqnarray}

	which, on its own, imposes equation of motion for the field $H$. 
	Using the solution (\ref{ccl}), we find  $H$ to be

	\begin{equation}
        H^{\lambda\mu\nu}= m \,\epsilon^{\lambda\mu\nu\rho}\, \left(\,
	\partial_\rho\, \phi - A_\rho \,\right) + g\, \partial^{[\,\lambda}\,
	{1\over \partial^2}\, J^{\mu\nu\,]}
	\label{hcl}
	\end{equation}
	
	Further integration of the $C$ field, using its solution
	(\ref{ccl}), gives

	\begin{eqnarray}
	 \int [D C] \,\delta\left[\, \partial_\mu \, C^{\mu\nu} 
	\,\right]\exp\left\{ -{1\over 4}
	\int d^4x\, C^{\mu\nu}\, C_{\mu\nu}\,\right\}=\nonumber\\  
	 \int [DA]\exp\left\{ -{1\over 4}\,
	\int d^4x\, F^*_{\mu\nu}(A) \,  F^{\ast\, \mu\nu}(A)
	\,\right\}\ ,
	\end{eqnarray} 
	
	while integration of the $H$ field, using (\ref{hcl}), gives 
	the final form of the dual generating functional as
	
	\begin{eqnarray}
	\fl \int [DA]\, [D H]\, \delta\left[\,
	H^{\lambda\mu\nu}   -  m\,\epsilon^{\lambda\mu\nu\rho}\,\left(\,
	\partial_\rho \phi - A_\rho\,\right)
	\,\right]\nonumber\\
	\fl \times\, \exp\left\{ \, 
	\int d^4x\,\left[\,   -  {1\over 4}\,  
	 F^*_{\mu\nu}(A) \,  F^{\ast\, \mu\nu}(A)
	  +{1\over 12}\,H^{\mu\nu\rho}\, H_{\mu\nu\rho} \,\right]
	 \,\right\}
	\nonumber\\  
	\fl =\int [DA][D\phi]  \,\exp\left\{  \int d^4x\,\left[\, 
	  -  {1\over 4} \, F_{\mu\nu}(A)\, F^{\mu\nu}(A)
	  +  {1\over 2}\, \left(\,m\,
	\left( \, \partial_\mu \, \phi -  A_\mu\,\right)
	+g\, \partial^{[\,\lambda}\,
	{1\over \partial^2}\, J^{\mu\nu\,]}\,\right)^2 \, \right]            
	\right\}\nonumber\\
	\label{finale}
	\end{eqnarray} 
	
	Equation (\ref{finale}) shows that the compensated, i.e. gauge 
	invariant, massive, KR theory is dual to a  compensated,  massive, 
	Proca theory (\ref{finale}). One would be tempted to believe that
	the presence of gauge invariance, both in compensated and massless case,
	implies the same dynamics. This way of thinking  
	suggests that (\ref{finale})   should be interpreted according to
	the scheme (\ref{m1}), leading to the following duality:
	
	\begin{eqnarray}
	B_{\mu\nu}  \longleftrightarrow   \phi \nonumber\\
	A_\mu  \longleftrightarrow  \phi_\mu\label{m4}
	\end{eqnarray}
	
        Scheme (\ref{m4}) implies that massive compensated KR field  is dual
	to a scalar field,{\it ergo}, it  has the same dynamical
	content of the massless case. The same reasoning, applied to the
	compensated Proca theory, tells that it should have the same dynamics
	as massless vector theory, i.e. two degrees of freedom.
	 It seems that there is an inconsistency in the spin content
	in (\ref{finale}) and  (\ref{hbfi}) if one follows the scheme
	(\ref{m4}), based on gauge symmetry. \\
	What happened to the spin jumping and
	what is the correct spin of the compensated, massive, KR field?
	 Is it determined by gauge invariance or by the presence of mass?
	 \\
        The  answer can only be found by counting the 
	number of physical (dynamical) degrees of freedom for the  
        theory given by the second order form of (\ref{hbfi}). 
	One could  do this  counting covariantly, a la' Fadeev--Popov, 
	but as already shown in \cite{pk} for the massless KR field,
	such counting turns out  not to be straightforward. Thus, 
	to avoid any ambiguity, we adopt a non--covariant description where the
	physical degrees of freedom are manifest.
	One start from the equation of motion for the compensated, massive,
	KR field 
	 
	 \begin{eqnarray}
	  \partial_\mu \, H^{\mu\nu\rho}(\,B\,)= m^2\left(\, B^{\nu\rho}-
	\partial^{[\,\nu}\, \phi^{\rho\,]}\,\right)\label{huno}\\
	 \partial_\nu \left(\, B^{\nu\rho}-
	\partial^{[\,\nu}\, \phi^{\rho\,]}\,\right)=0 \label{hdue}
	\end{eqnarray}
	
	where, (\ref{hdue}) is a consistency condition following
	from (\ref{huno}). Equation (\ref{hdue}) allows us to solve the
	compensator  $\phi_\mu$ in terms of $ B_{\lambda\mu}$ as
	
	\begin{equation}
	\phi_\mu= \partial^\lambda\, {1\over \partial^2}\, B_{\lambda\mu}
	\label{fit}
	\end{equation}
	
	Inserting (\ref{fit}) in (\ref{hdue}) one obtains a {\it non--local,
	gauge invariant,} field equation for $B$:
	
	\begin{equation}
	\left(\,\partial^2 -m^2 \,\right)\,\partial_\mu\, {1\over \partial^2} 
	\, H^{\mu\nu\rho}(\,B\,)= 0
	\label{htre}
	\end{equation}
	
	To extract the physical degrees of freedom  contained in (\ref{htre}),
        we write separately its $(\, i\,0\,)$ and $(\, i\,j\,)$ 
	components. The components $(\, i\, 0\,)$ of (\ref{htre}) give the
	equation of motion for the transverse vector field 
	$ C^i\equiv \partial_j\, H^{j i0}$ as
	
	\begin{equation}
	\left(\, \partial^2 - m^2 \,\right)\,{1\over \partial^2 }\,\vec C=0
	\label{vt}
	\end{equation}
	
	Furthermore, the $(\, i\, j\,)$ components of (\ref{htre}) and
	(\ref{vt}) give additional equation of motion for the ``scalar''
	component as
	
	\begin{equation}
	\left(\, \partial^2 - m^2 \,\right)\, \epsilon^{ijk}\, H_{ijk}\left(\,
	B\,\right) =0
	\label{scal}
	\end{equation}
	
	Equations
	(\ref{vt}) and (\ref{scal}) describe the dynamics of the compensated,
	massive, KR field and show that 
	there are {\it three} dynamical degrees of freedom (spin one) instead of
	a single (spin zero) scalar component.  This is the answer to the 
	previously raised question:
        massive, compensated, KR field describes a spin one field. 
        From the dualization result (\ref{finale}), one concludes that 
        also massive, compensated, Proca field has  three degrees of freedom 
        \footnote{ This conclusion can be independently confirmed by direct
        analysis of dynamical content in the same way as for KR field. }
        and not two as one would naively believe. The conclusion is 
        that in general,  massive, compensated, gauge theory {\it is not } 
        dynamically equivalent to a massless gauge theory, in spite of the gauge
        invariance of both theories. As a consequence one has to follow the 
        dualization scheme (\ref{m2}) for massive $p$--forms. This brings us to 
        the non-trivial conclusion that the spin jumping is caused by the 
        {\it sole} presence  of the mass term.  Gauge symmetry is not a 
        necessary ingredient for this effect to take place.\\
        Let us stress the difference between compensated and non--compensated 
        massive theories. In the absence of compensator, the mass term is not
        gauge invariant and the
        consistency condition $\partial_\mu\, B^{\mu\nu}=0$ relates the 
        components of KR  field so that $C^i=\partial^2\, B^{0i}_T$. 
        Then, the equations of motion (\ref{vt}),(\ref{scal}) become
	
	\begin{eqnarray} 
	 \left(\, \partial^2 - m^2 \,\right)\, B^{i0}_T=0  \label{et}\\
	 \left(\, \partial^2 - m^2 \,\right)\, \epsilon^{ijk}\, H_{ijk}\left(\,
	B\,\right)  =0\label{bl}
	\end{eqnarray} 
	
	showing, again, three dynamical degrees of freedom as well as the gauge
	non invariance of (\ref{et}), which is explicitly dependent on the 
	transverse components of the vector $B^{i0}$.
	On the other hand,  in the massless case the 
	covariant  equations of motion (\ref{htre}) combine to give the 
	propagation of the scalar degree of freedom
	
	\begin{equation}
	\partial^2\, \epsilon^{ijk}\, H_{ijk}\,\left(\,B\,\right) =0 
	\label{massless}
	\end{equation}
	
	In this letter we have investigated massless, massive, and
	{\it massive, Stueckelberg compensated} phases of the
	Kalb-Ramond field theory. We have extended the dualization procedure
	of \cite{aass} to the latter two phases and shown their duality
	to ordinary and compensated Proca theories. The result of duality
	procedure shows the effect of spin jumping for the KR field.
	While in massless and ordinary massive phase the spin content is
	clearly exhibited, in the compensated phase it is not so obvious
	and we have shown that this phase also describes a spin one field.
	The advantage of the compensated phase is that one can consider
	simultaneously mass and gauge symmetry and study their effects on spin
	jumping. The result shows that the spin jumping is caused only by the 
	presence or absence of mass.

	\end{document}